\documentclass[pra,twocolumn,showpacs,preprintnumbers,amsmath,amssymb]{revtex4-1}
\usepackage{graphicx}
\usepackage{amsmath}
\usepackage{MnSymbol}
\usepackage{mathrsfs}
\usepackage{colortbl}
\usepackage[table]{xcolor}
\usepackage{tabularx}


\begin{document}
\title{Were there two forms of \textit{Stegosaurus}?}

\author{Robert P. Cameron$^\ast$, John A. Cameron and Stephen M. Barnett}
\affiliation{\mbox{School of Physics and Astronomy, University of Glasgow, Glasgow G12 8QQ, United Kingdom} \\
$\ast$r.cameron.2@research.gla.ac.uk}


\begin{abstract}
\noindent We recognise that \textit{Stegosaurus} exhibited exterior chirality and could, therefore, have assumed either of two distinct, mirror-image forms. Our preliminary investigations suggest that both existed. \textit{Stegosaurus}'s exterior chirality raises new questions such as the validity of well-known exhibits whilst offering new insights into long-standing questions such as the function of the plates. We inform our discussions throughout with examples of modern-day animals that exhibit exterior chirality.
\end{abstract}
\maketitle


\noindent\textit{Stegosaurus} was discovered well over a century ago, during the infamous Bone Wars \cite{Marsh77}. It remains one of the most readily recognisable genera of dinosaurs, owing to its distinctive plates. The precise arrangement of these plates and, indeed, their function, have been the subject of much debate, however \cite{Lull1910a, Lull1910b, Gilmore14, Farlow76,Buffrenil86,Czerkas87,Carpenter98a, Carpenter98b}. The currently favoured reconstruction sees them stand upright in two staggered rows that run the length of the beast, as evidenced for \textit{Stegosaurus stenops} in particular by the holotype of this species \cite{Marsh87, Gilmore14, Gilmore15, Gilmore18} and other articulated skeletons besides \cite{Carpenter98a}. The present paper is concerned with a subtle geometrical property inherent to \textit{Stegosaurus}'s plates and its implications. 

The word \textit{chiral} was introduced to describe any geometrical figure or group of points that cannot be brought into coincidence with its mirror image, thus possessing a sense of handedness \cite{Kelvin94}. It derives from the Greek word for hand; $\chi\epsilon \acute{\iota}\rho\alpha$ \cite{Electra15}. Chirality pervades the natural world, from the enigmatic preferences of fundamental physical forces \cite{Lee56, Wu57} to the arms of spiral galaxies \cite{Kondepudi01}. It is of particular importance to life, as the molecules that comprise living things are invariably chiral and their chirality is crucial to their biological function \cite{Gardner90, Lough02}.

In spite of this, the overwhelming majority of living things possess rather symmetrical, achiral exterior forms, leaving but a small handful of living things that instead boast exterior chirality, outwardly defying mirror symmetry \cite{Neville76, Gardner90, Lough02, website15a}. The shell of the escargot snail \textit{Helix pomatia}, for example, is manifestly chiral as it may exhibit either a left- or a right-handed twisting form, these being non-superposable mirror images of each other, as seen in FIG. 1.\newline\newline

\begin{figure}[h!]
\centering
\includegraphics[width=\linewidth]{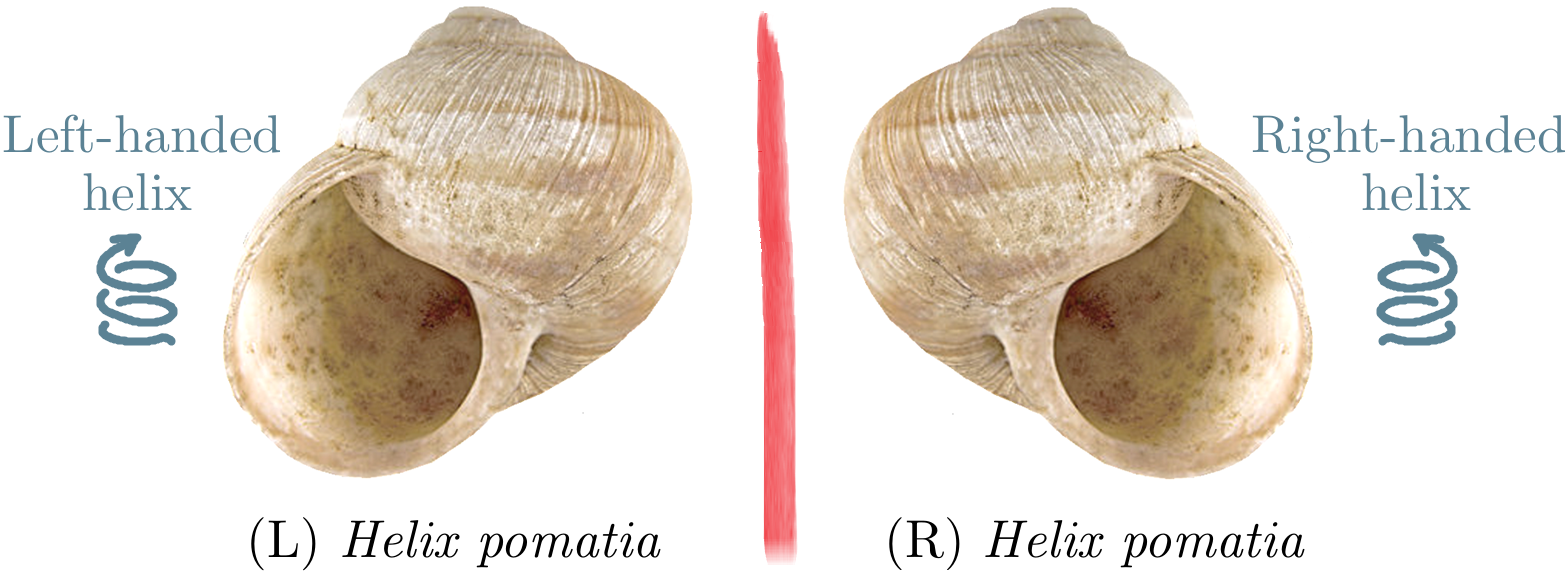}\\
\caption{\small \textit{Helix pomatia}'s shell differs from its mirror image and is thus chiral \cite{FIGURE1CAP}.} 
\label{Fig1}
\end{figure}

\begin{figure}[h!]
\centering
\includegraphics[width=\linewidth]{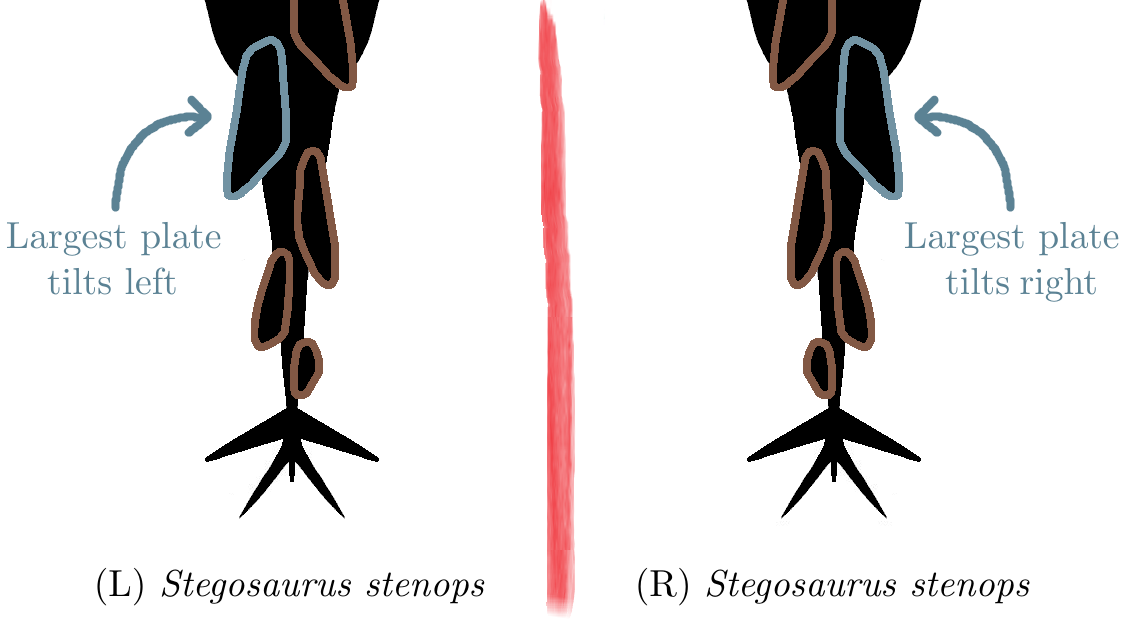}\\
\caption{\small \textit{Stegosaurus}'s plates differ from their mirror image and are thus chiral.} 
\label{Fig2}
\end{figure}

It seems that \textit{Stegosaurus} also exhibited exterior chirality. The currently favoured arrangement of \textit{Stegosaurus}'s plates is clearly distinct from its mirror image, as seen in FIG. 2. Moreover, some of the plates display subtle exterior chirality individually \cite{Lull1910a, Gilmore14} and no two plates of the same size and shape have ever been found for the same specimen \cite{Lull1910a, Gilmore14, Czerkas87}. In spite of its simplicity, this observation, that \textit{Stegosaurus} exhibited exterior chirality, does not appear to have been made explicitly before and, more importantly, its implications do not appear to have been recognised. \textit{Stegosaurus}'s plates and their arrangement have certainly been described as being ``asymmetrical'' \cite{Lull1910a, Gilmore14, Czerkas87}, but this is not quite synonymous with being chiral \cite{Barron04}. Chirality carries with it the connotation that the mirror-image form is also viable, whether it exists naturally or not. In this sense, chirality is as much about symmetry as the lack thereof. The term \textit{dissymmetry}, introduced by Pasteur, comes to mind in this regard. As Barron wrote ``Dissymmetric figures are not necessarily \textit{asymmetric}, that is devoid of all symmetry elements ... However, dissymmetry excludes improper rotation axes, that is centres of inversion, reflection planes and rotation-reflection axes.'' \cite{Barron04}.

Just as one can distinguish between two distinct, mirror-image forms of snail on the basis of whether the shell twists in a left- or a right-handed manner, we distinguish between two distinct, mirror-image forms of \textit{Stegosaurus}, depending on whether the largest plate, located over the base of the tail for \textit{Stegosaurus stenops} at least \cite{Gilmore14,Gilmore15,Carpenter98a}, tilts to the left or to the right as seen when looking down upon a specimen. We designate these (L) \textit{Stegosaurus} and (R) \textit{Stegosaurus}, respectively. 

A variety of interesting questions follow. Our goal here is not to try and answer these definitively but rather to highlight their existence for the benefit of future discussions. Many of these ideas may be subtly interrelated, of course.

In what proportion did (L) \textit{Stegosaurus} and (R) \textit{Stegosaurus} specimens exist, for a given species, gender and location? Perhaps one form was genetically favoured with the other being found only rarely due to mutation, as in the case of \textit{Helix pomatia} which is found predominantly in its right-handed form, with perhaps 1 in 20,000 specimens having a left-handed form instead \cite{Lough02}. Another possibility is that (L) \textit{Stegosaurus} and (R) \textit{Stegosaurus} existed in equal proportion, as in the case of the Portugese man-o'-war \textit{Physalia physalis} which is equally likely to be born with its sail pointing to the left or to the right, perhaps to reduce the risk of all offspring being washed ashore \cite{Neville76}. For the species \textit{Stegosaurus stenops} (gender unknown) in Colorado, the holotype \cite{Marsh87,Gilmore14,Gilmore15,Gilmore18,Czerkas87} appears to have belonged to an (R) specimen whereas the ``Small skeleton'' \cite{Carpenter98a} may, perhaps, have belonged to an (L) specimen, an assignment that seems to be in accord with Carpenter's own figures at least \cite{Carpenter98a}. Tentatively then, it seems we have a positive answer to our question: yes, there \textit{were} two forms of \textit{Stegosaurus}. The interpretation of these remains is far from unambiguous, however, and there may yet be a subtle twist besides: the illustrations of the holotype seen in \cite{Gilmore14,Gilmore15,Czerkas87} appear to derive from lithographic plates and could, therefore, indicate the wrong chirality, as lithographic techniques often yield reversed images \cite{Lough02}. In an attempt to clarify this issue, we scrutinised the wonderful monograph by Gilmore \cite{Gilmore14}, who prepared the holotype for exhibition at the United States National Museum in Washington D.C., now the National Museum of Natural History. There we found an explicit statement by him that the specimen was indeed found lying ``on its left side'', which is in accord with our suggestion of (R) chirality. Gilmore justifies this statement, however, by quoting a letter written by Welch (who discovered the specimen) to Marsh (who announced the discovery \cite{Marsh87}) in turn as ``the animal lay on its left side and up against the bank of our river bed, bringing its left [right] hip the highest, the right [left] hip and some bones having slid downhill toward the bottom of the bed''. Without Gilmore's corrections, in square brackets, this description is, in fact, ambiguous with regards to the chirality of the specimen. It seems likely that Gilmore and, indeed, Lucas (who examined the holotype before Gilmore \cite{Gilmore14}) would have had additional information that guided these choices. Further digging on this matter is required in order to be certain, however.

Did \textit{Stegosaurus} chirality vary between species? For example, the American lightning whelk \textit{Busycon contrarium} is found predominantly with its shell twisting in a left-handed manner whereas the American channelled whelk \textit{Busycon canaliculatum} is found predominantly with its shell twisting in a right-handed manner \cite{Lough02}. One of the most complete \textit{Stegosaurus} skeletons in the world was initially ascribed to the species \textit{Stegosaurus armatus} and named ``Sarah'', after the daughter of one John Ed Anderson; the owner of the ``Red Canyon Ranch'' in Wyoming where ``Sarah'' (the \textit{Stegosauru}s, gender unknown)  was found \cite{Siber09}. ``Sarah'' appears to have been an (R) specimen \cite{Siber09}, matching the (R) chirality suggested above for the \textit{Stegosaurus stenops} holotype . ``Sarah'' has recently been ascribed to the species \textit{Stegosaurus stenops} and renamed ``Sophie'', after the daughter of one Jeremy Herrmann; a generous donor to the Natural History Museum in London \cite{website15b}.

\begin{figure}[h!]
\centering
\includegraphics[width=\linewidth]{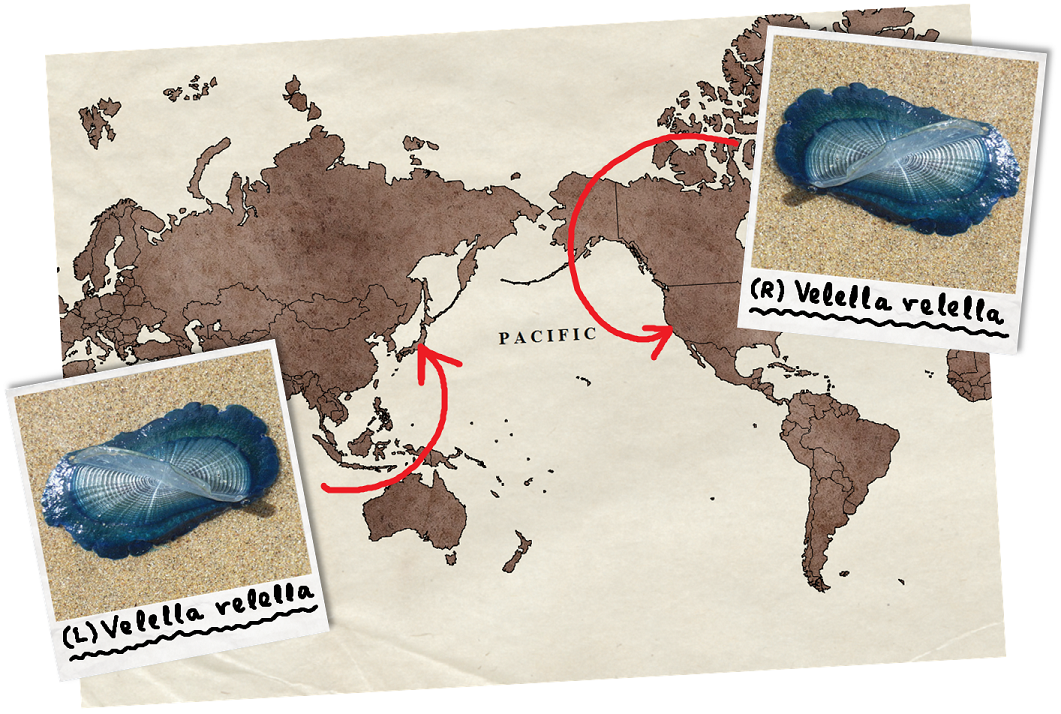}\\
\caption{\small Exterior chirality can yield information about migration, as in the case of \textit{Velella velella}, specimens of which tack in different directions depending upon the orientation of their sails relative to their bodies \cite{Gardner90,FIGURE3CAP}.} 
\label{Fig3}
\end{figure}

Did \textit{Stegosaurus} chirality vary between genders? For example, the \textit{Papaya} flower has petals that twist clockwise for females but anticlockwise for males \cite{Lough02}. This is a rather timely question, as it has been suggested recently that the plates of \textit{Hesperosaurus mjosi}, closely related to the \textit{Stegosaurus} genus, differed in size and shape between males and females. If correct, this would constitute one of the few examples known of sexual dimorphism in dinosaurs \cite{Saitta15}. It is conceivable also that the chirality of the plates was important for the mechanics of \textit{Stegosaurus} mating, which remain very poorly understood. An example of note here is that of the South American tooth-carp \textit{Anableps anableps}, both genders of which possess chiral sex organs such that any given specimen is only capable of mating with half of the population of the opposite gender \cite{Neville76}. 

Did \textit{Stegosaurus} chirality vary with geography? For example, the marine polyp \textit{Velella velella} is found on the Japanese side of the Pacific with its sail directed one way and the North American side of the Pacific with its sail directed the other way, as depicted in FIG. 3. It is believed that these distinct, mirror-image forms occur in equal proportion in the middle of the Pacific and that the wind separates them \cite{Gardner90}. Analogously, geographical variations in \textit{Stegosaurus} chirality, if found, could aid in our understanding of \textit{Stegosaurus} migration: it was thought until very recently that \textit{Stegosaurus} lived exclusively in North America; a paradigm that shifted with the unexpected discovery of a skeleton in Portugal ascribed to the species \textit{Stegosaurus ungulates} \cite{Escaso07}. Unfortunately, this skeleton is insufficiently complete for us to speculate as to chirality of the specimen for comparison with the chiralities suggested above of \textit{Stegosaurus} specimens from North America. 

\begin{figure}[h!]
\centering
\includegraphics[width=0.8\linewidth]{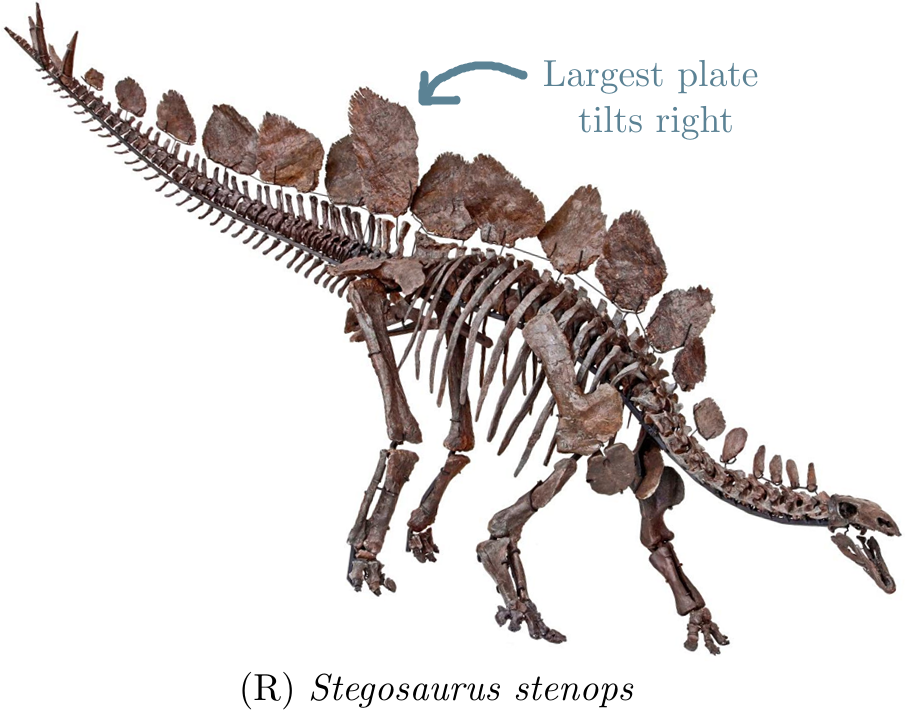}\\
\caption{\small This (R) \textit{Stegosaurus stenops} skeleton on display at the Natural History Museum in London \cite{website15b} seems to be a faithful representation of the original specimen. The validity of other \textit{Stegosaurus} exhibits is less clear \cite{FIGURE4CAP}.} 
\label{Fig4}
\end{figure}

A particularly interesting question is whether or not skeletal reconstructions are truly faithful representations of the animals to which the bones originally belonged. It could be, for example, that a skeleton unwittingly mounted as an (R) \textit{Stegosaurus} originally belonged to an (L) \textit{Stegosaurus} or \textit{vice-versa} and is thus a subtle misrepresentation. Gilmore exhibited the remains of the \textit{Stegosaurus stenops} holotype in the same layout as they were supposedly found \cite{Gilmore14}, with (R) chirality. He also installed a large mirror to make the underside of the skeleton visible \cite{Gilmore14}. The creature seen in this mirror would have had the opposite chirality, of course, to that of the exhibit itself. We can say with certainty then that one of the two views of the \textit{Stegosaurus stenops} holotype on offer constituted a faithful representation of the original specimen whilst the other did not. ``Sophie'' is also mounted with (R) chirality at present, as seen in FIG. 4. This matches the (R) chirality suggested above of the original specimen. Analogous questions can be raised with regards to depictions of \textit{Stegosaurus} in popular culture. The specimens shown in the 1997 film ``The Lost World: Jurassic Park'', for example, appear once more to be of (R) chirality.

Three main hypotheses for the function of \textit{Stegosaurus}'s plates have been put forward over the years. The chirality of the plates offers new insights here. (i) It was first suggested that the plates served as a form of armour \cite{Marsh77}. This is now thought to be unlikely, however, as the plates were seemingly too fragile and ill-placed \cite{Carpenter98a}. Indeed, it is known that \textit{Stegosaurus} engaged in fierce battles with \textit{Allosaurus}, during which \textit{Allosaurus} would occasionally bite off sizeable portions of \textit{Stegosaurus}'s plates with apparent ease \cite{Carpenter05}. The chiral arrangement of the plates may be regarded as further evidence against their use in a mechanically protective role, as it would have seen a specimen slightly more vulnerable to attack from one side than the other with no obvious benefit. (ii) Another possibility is that the plates served to regulate body temperature \cite{Farlow76,Buffrenil86}. The suggestion in particular that they acted as ``forced convection fins'' would seem to necessitate a staggered or ``interrupted'' pattern \cite{Farlow76}. That is, the chiral arrangement of the plates but with no obvious reason to prefer (L) \textit{Stegosaurus} over (R) \textit{Stegosaurus} or \textit{vice-versa}. These ideas are now thought to be unlikely as similar dinosaurs with analogous but smaller dermal features such as \textit{Kentrosaurus} apparently flourished in comparable climates \cite{Carpenter98a}. (iii) The most popular idea at present is that the plates served as display structures, perhaps to ward off potential predators, to aid in identification or as a means of attracting mates \cite{Gilmore14, Carpenter98a, Carpenter98b}. The high degree of vascularisation evident in the plates \cite{Buffrenil86} has led in particular to claims that they could ``blush'' so as to embellish their appearance \cite{Carpenter98a,Carpenter98b}. The chirality of the plates makes the idea that they served as display structures seem all the more plausible to us, for it is integral to their appearance. We note in particular that two staggered rows of plates gives a more substantial lateral profile than would two parallel rows of plates, for example, as the latter yields visible gaps where the former has none. We are reminded here of the eel \textit{Leptocephalus diptychus}, which is seen in its post-larval stage to possess four spots on one side of its body and three on the other. These spots alternate in position such that all seven can be seen from either side of the eel, which is otherwise transparent \cite{Neville76}. The chiral arrangement of the plates may be thought of then as a means of giving a body-length sail of maximised apparent area whilst permitting freedom of movement. If we suppose that the purpose of this sail was to ward off potential predators, we see no obvious reason to prefer (L) \textit{Stegosaurus} over (R) \textit{Stegosaurus} or \textit{vice-versa}, as the size alone of the sail would likely be its most important characteristic. More subtle possibilities arise if we suppose that the sail aided in identification or as a means of attracting mates. A given specimen would have appeared somewhat different when viewed from the left or from the right and may, therefore, have preferred to display one side over the other, clearly distinguishing its head and tail ends in the process. Such chiral behaviour would likely need to be hard-wired given \textit{Stegosaurus}'s apparently limited mental capacity and so would seem to demand consistent exterior chirality between similar specimens. Chiral behaviour is exhibited by kangaroos and wallabies, for example, which show a preference for their left forelimbs \cite{Giljov15}. We recognise another possibility here. It seems likely that \textit{Stegosaurus} would have spent much of its time grazing in amongst luscious vegetation so as to maintain its enormous size, all the while hoping to avoid fierce antagonists like \textit{Allosaurus}. \textit{Stegosaurus}'s angular and slightly curved plates are not entirely unlike the fronds of a fern, for example. It is at least conceivable then that the plates functioned as a form of camouflage, with their chirality serving simply to give a more convincing appearance. We are reminded here of various salamanders, frogs, butterflies and snakes that exhibit chiral patternings as a form of camouflage \cite{Neville76}. In this role there is again no obvious reason to prefer (L) \textit{Stegosaurus} over (R) \textit{Stegosaurus} or \textit{vice-versa}. 

\begin{figure}[h!]
\centering
\includegraphics[width=\linewidth]{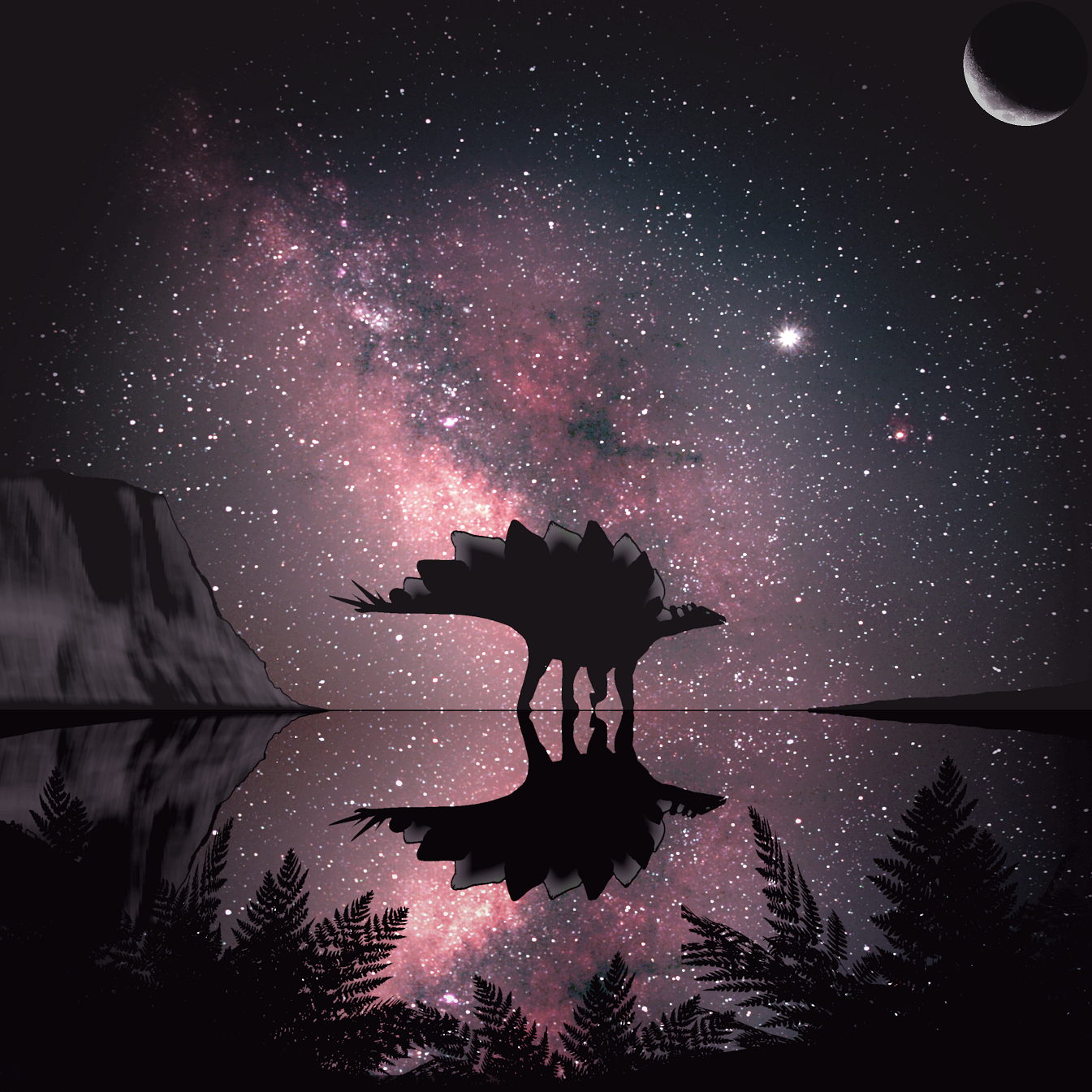}\\
\caption{\small An illustration of \textit{Stegosaurus}'s exterior chirality.} 
\label{Fig5}
\end{figure}

\textit{Stegosaurus}'s exterior chirality may be of particular interest to evolutionary biologists. It is one of the earliest examples known to us of exterior chirality in a living thing. Moreover, exterior chirality is especially rare amongst present-day reptiles and birds \cite{Lough02}.

Tacit in the above is a focus upon adolescent and adult specimens. It has been suggested that the ``asymmetry'' of the plates' arrangement may have been absent from juveniles, appearing only in later life, the dermal spikes of the rhinoceros iguana \textit{Cyclura cornuta} having been cited by way of example \cite{Czerkas87}. Remains found more recently suggest that juveniles may not have had plates at all \cite{Galton10}. We are reminded here of flatfish such as the starry flounder \textit{Platichthys stellatus}. These too are born as seemingly symmetrical creatures only to develop exterior chirality thereafter \cite{Policansky82,Gardner90}. If it is indeed the case that the principal role of the plates was for display in courtship, then it is even possible that only males had well-developed plates. This would go some way, of course, towards demystifying the mechanics of \textit{Stegosaurus} mating.

Progress would be greatly advanced, of course, by a more expansive survey of \textit{Stegosaurus}'s exterior chirality. At first glance this would seem to require that more articulated skeletons be discovered. Certainly chirality should be in the minds of those lucky enough to make such finds. An expansive survey may already be viable, however, given the current fossil record. It may be possible to distinguish between (L) and (R) specimens by considering the size and shape of individual plates alone \cite{Lull1910a, Gilmore14}, which would negate the need for articulated skeletons. Partial skeletons or even single plates may also be of value in this exercise if they bear marks from predation. A predator, such as \textit{Allosaurus}, would have bitten into plates closest to it and if the orientation of the bite can be determined from the remaining teeth marks in damaged \textit{Stegosaurus} plates \cite{Carpenter05}, then the arrangement of the plates and hence the specimen's exterior chirality should follow. 

It is our hope that the present paper will inspire explicit consideration of exterior chirality in future investigations of \textit{Stegosaurus} and, indeed, other such dinosaurs. 

This work was supported by the Engineering and Physical Sciences Research Council under grants EP/M004694/1 and EP/I012451/1.  We thank Revinder Chahal at the Natural History Museum in London for her swift and helpful correspondences.



\end{document}